\documentstyle[preprint,prl,aps]{revtex}

\begin{document}
%
%
\def\jcomp#1#2#3{J. Comput.\ Phys.\ {\bf #1}, #2 (#3)}
\def\jpa#1#2#3{J. Phys.\ A {\bf #1}, #2 (#3)}
\def\jpc#1#2#3{J. Phys.\ C {\bf #1}, #2 (#3)}
\def\jsc#1#2#3{J. Supercond.\ {\bf #1}, #2 (#3)}
\def\nat#1#2#3{Nature {\bf #1}, #2 (#3)}
\def\phb#1#2#3{Physica B {\bf #1}, #2 (#3)}
\def\phc#1#2#3{Physica C {\bf #1}, #2 (#3)}
\def\pla#1#2#3{Phys.\ Lett.\ A {\bf #1}, #2 (#3)}
\def\pr#1#2#3{Phys.\ Rev.\ {\bf #1}, #2 (#3)}
\def\prb#1#2#3{Phys.\ Rev.\ B {\bf #1}, #2 (#3)}
\def\prl#1#2#3{Phys.\ Rev.\ Lett.\ {\bf #1}, #2 (#3)}
\def\pssb#1#2#3{Phys.\ Status Solidi (b) {\bf #1}, #2 (#3)}
\def\rmp#1#2#3{Rev.\ Mod.\ Phys.\ {\bf #1}, #2 (#3)}
\def\zpb#1#2#3{Z.\ Phys.\ B {\bf #1}, #2 (#3)}
\def\pdp#1#2{\frac{\partial}{\partial #2} #1}
\def\pdw#1#2{\frac{\partial #1}{\partial #2}}
\font\testfont=msam10
\newcommand{\Bullet}{\mbox{\raisebox{-1.0mm}{\huge $\bullet$}}}
\newcommand{\circledbullet}{{{\large $\bullet$}%
\hspace{-0.82em}$\bigcirc$}}
\newcommand{\solidiamond}{{\testfont \char7}}
\newcommand{\eq}[1]{Eq.~(\ref{#1})}
\newcommand{\fig}[1]{Fig.~\ref{#1}}
\newcommand{\figs}[3]{Figs.~\ref{#1}(#2) and (#3)}
\newcommand{\mc}[2]{\multicolumn{#1}{#2}}
\newcommand{\tab}[1]{Table~\ref{#1}}
\newcommand{\hypen}{{\mbox{-}}}
\newcommand{\half}{$\frac{1}{2}$}
\newcommand{\quatro}{$\frac{1}{4}$}
\newcommand{\pioa}{$\frac{\pi}{a}$}
\newcommand{\piob}{$\frac{\pi}{b}$}
\newcommand{\pioc}{$\frac{\pi}{c}$}
\newcommand{\aaocc}{$\frac{a^2}{c^2}$}
\newcommand{\eg}{{\it e.g.}}
\newcommand{\etal}{{\it et al.}}
\newcommand{\ie}{{\it i.e.}}
\newcommand{\abi}{{\it ab initio}\,}
\newcommand{\Abi}{{\it Ab initio}\,}
\newcommand{\Tc}{{\it T}$_c$}
\newcommand{\htc}{{high-\Tc}\,}
\newcommand{\Htc}{{High-\Tc}\,}
\newcommand{\lda}{local density approximation}
\newcommand{\ldf}{local density functional}
\newcommand{\dft}{density functional theory}
\newcommand{\dritz}{DIIS-{\em ritzit}}
\newcommand{\pcg}{preconditioned conjugate gradient}
\newcommand{\pp}{pseudopotential}
\newcommand{\pppw}{pseudopotential plane-wave}
\newcommand{\k}{{\bf k}}
\newcommand{\ef}{$E_F$}
\newcommand{\au}[1]{#1 \mbox{a.u.}}
\newcommand{\ev}[1]{$#1$ eV}
\newcommand{\ry}[1]{#1 \mbox{Ry}}
\newcommand{\nef}{$N(E_F)$}
\newcommand{\dos}[1]{#1 states/(eV unit-cell)}
\newcommand{\cudos}[1]{{$#1$~states/(eV Cu)}}
\newcommand{\zcord}[1]{$z_{\rm\scriptscriptstyle #1}$}
\newcommand{\sm}{\mbox{-}}
\newcommand{\ket}{\rangle}
\newcommand{\bra}{\langle}
\newcommand{\dpsi}[1]{{|\delta \psi^{(#1)}\ket}}
\newcommand{\vpsi}[1]{{|\psi^{(#1)}\ket}}
\newcommand{\e}[1]{\varepsilon^{(#1)}}
\newcommand{\cuo}{{CuO$_2$}}
\newcommand{\cco}{{CaCuO$_2$}}
\newcommand{\csco}[2]{{Ca$_{#1}$Sr$_{#2}$CuO$_2$}}
\newcommand{\dcsco}[2]{{(Ca$_{1\sm#1}$Sr$_{#1}$)$_{1\sm#2}$CuO$_2$}}
\newcommand{\lco}{{La$_2$CuO$_4$}}
\newcommand{\lbco}[2]{{La$_{#1}$Ba$_{#2}$CuO$_4$}}
\newcommand{\lmco}[2]{{La$_{#1}M_{#2}$CuO$_4$}}
\newcommand{\lsco}[2]{{La$_{#1}$Sr$_{#2}$CuO$_4$}}
\newcommand{\sybco}{{YBa$_2$Cu$_3$O$_7$}}
\newcommand{\ybco}[2]{{YBa$_2$Cu$_{#1}$O$_{#2}$}}
\newcommand{\bscco}[2]{{Bi$_2$Sr$_2$CaCu$_{#1}$O$_{#2}$}}
\newcommand{\bi}[3]{{Bi$_2$Sr$_2$Ca$_{#1}$Cu$_{#2}$O$_{#3}$}}
\newcommand{\tl}[3]{{Tl$_2$Ba$_2$Ca$_{#1}$Cu$_{#2}$O$_{#3}$}}
\newcommand{\xy}{{x^2-y^2}}
\newcommand{\yz}{{y^2-z^2}}
\newcommand{\zx}{{z^2-x^2}}
\newcommand{\zr}{{3z^2-r^2}}

%
%
\draft

\title{\Abi\ Pseudopotential Plane-wave Calculations of
the Electronic Structure of \sybco}

\author{Hanchul Kim and Jisoon Ihm}
\address{Department of Physics and Center for Theoretical Physics,\\
         Seoul National University, Seoul, 151-742, Korea}

\date{September 15, 1994}
\maketitle

\begin{abstract}

    We present an \abi\ \pp\ \ldf\ calculation for stoichiometric
\htc\ cuprate \sybco\  using the plane-wave
basis set.  We have overcome well-known difficulties in applying
\pp\ methods to first-row elements, transition metals, and
rare-earth materials by carefully generating norm-conserving \pp s
with excellent transferability and employing an extremely efficient
iterative diagonalization scheme optimized for our purpose.  The
self-consistent band structures, the total and site-projected
densities of states, the partial charges and their 
symmetry-decompositions, and some characteristic charge densities 
near \ef\ are presented.  We compare our results with
various existing (F)LAPW and (F)LMTO calculations and establish
that the \abi\ \pp\ method is competitive with other methods in
studying the electronic structure of such complicated materials as
\htc\ cuprates.
\end{abstract}
\bigskip
\pagebreak
\section{Introduction}
%
%
The discovery of superconductivity above 30K in \lbco{2-x}{x} by
Bednorz and M\"{u}ller~\cite{Bednorz} has stimulated intensive
studies on various cuprates as well as the mechanism
underlying the phenomenon of \htc\ superconductivity.  A wide
variety of speculations about its microscopic origin have been
raised with the sole consensus of the apparently important role of
the
\cuo\ layer (and, of course, the pairing of charge carriers) which is
the basic building block of \htc\ cuprates.  Even for the 
{\em normal} (above \Tc) state of the \htc\ cuprates, the controversy 
as to whether it is a Fermi liquid or some other novel states has not
been settled.  The reason for the complication is that many of the
cuprate superconductors have sister compounds which are strongly
correlated insulating antiferromagnets.  
Among novel points of view explaining anomaly in normal 
states of \htc\ cuprates are the Luttinger liquid 
theory, the marginal Fermi liquid theory, the van Hove scenario, 
the spin fluctuation theory, the extended Hubbard model, the 
t-{\it J}\, model, and the anyon theory, to mention a few.  On
the other side of such efforts, there has been a more conventional
approach dealing with the full Hamiltonian and the realistic atomic
structure in the belief of the Fermi liquid theory; the first
principles density functional theory~\cite{Hohenberg} with
the local density approximation (LDA)~\cite{Kohn} for the
exchange-correlation energy is such an example.  In spite of its
shortcomings in the magnetic insulating phase, the various
state-of-the-art electronic structure calculations have proven to
be successful in predicting from first principles a number of
properties of metallic cuprates such as the lattice constant,
atomic positions within the unit cell, phonon frequencies,
structural instabilities, electric field gradients, and most
importantly, the detailed energy band structure and the Fermi
surface (FS)~\cite{Pickett89}. 
Such calculations show the existence of a half-filled 
$pd\sigma$-antibonding band and the resulting cylindrical
FS per \cuo\ plane as the distinctive electronic feature of the 
\htc\ cuprates~\cite{Mattheiss93}. 

        Most of the existing electronic structures for \htc\
cuprates have been calculated with use of the all-electron 
localized basis methods (LMTO, FLMTO, LAPW, or FLAPW) or 
the \pp\  mixed basis method.
Here, we want to emphasize the merits of the plane-wave basis set, 
which above-mentioned methods do
not share.  The mathematical formulation and the implementation of
the numerical codes are particularly simple with the plane-wave
basis set.  The accuracy of calculation is easily controlled
by varying the kinetic energy cutoff.  Plane-waves form a basis set
independent of the ionic positions, allowing for an unbiased
uniform description of the system and a much simpler calculation of
quantum mechanical (Hellman-Feynman) forces.    Furthermore, plane
waves enable us to use the fast Fourier transform so that the
iterative diagonalization
methods~\cite{Davidson,Liu,Wood,Martins88,Martins91}
or the \abi\ molecular dynamics~\cite{Car,Teter,Payne} can be
applied efficiently.  In conjunction with the \pp\
formulation~\cite{Hamann,Bachelet} where the rapid oscillations of
valence wavefunctions in the core region are systematically
smoothed out with their norms kept conserved, the plane-wave basis
sets are quite successful in the calculation of the electronic
structure of most semiconductors and some metals.  However, until
recently it was believed that materials including first-row
elements, transition-metal atoms, and rare-earth elements are not
suitable for the application of the \pppw\ method.
  With the development of efficient methods for
diagonalizing large Hamiltonian matrices represented in a
plane-wave basis set~\cite{Davidson,Liu,Wood,Martins88,%
Martins91,Car,Teter,Payne}, together with relatively smooth
\pp s~\cite{Troullier,Rappe},
it is now possible to perform the electronic structure calculations
for systems containing such atoms.  

	In this paper, we present an \abi\ \pp\ density-functional
calculation of \sybco\ using the plane-wave basis set. 
    \ybco{3}{7-\delta}  $(0\leq\delta\leq 1)$ is perhaps the most
extensively studied compound among the cuprate superconductors. 
A unique
feature of this compound is that the three copper atoms in
the primitive cell play two different roles.  Two copper atoms
(Cu2) participate in two conduction layers of \cuo\ separated by
yttrium atoms.  The third copper atom (Cu1) has an unusual
coordination with oxygen atoms, forming a one-dimensional CuO chain
structure along the $b$ direction.  
The stoichiometric compound \sybco\ is particularly interesting 
from the standpoint of band theorists.  It crystallizes in the
relatively simple structure (with 13 atoms in the orthorhombic 
primitive cell) allowing for the computation
with a less severe load compared with Bi- or Tl-based cuprate
superconductors.  More importantly, it is a metal instead of an
antiferromagnetic insulator in its normal state in contrast to
other stoichiometric cuprates like \lco.  Aside from the problem in 
handling the strong correlation among electrons, the \ldf\ (LDF) 
calculation is applicable to this system.  Furthermore,
clean surfaces are available in this compound so that high
resolution spectroscopic experiments can be performed and the
theoretical results can be compared directly with experiments.  The
quasiparticle band near the Fermi level and the FS
topology from the ARPES~\cite{Kampuzano}, ACAR~\cite{Haghighi}, and
dHvA~\cite{Fowler} experiments are in good accordance with those
obtained from the existing LDF calculations~\cite{Massidda87,%
JYu87b,Krakauer,Massidda90,Pickett90,Schwarz,Andersen,Hiroi,%
Mazin92b}.
This fact is sometimes regarded as supporting evidence for the
Fermi liquid picture of the normal state of \htc\ cuprates though
the issue is still controversial.
  Since the main purpose of the
present work is to show the feasibility of the \abi\ \pppw\ method
for studying \htc\ materials in
comparison with other methods and since the present paper is the
first such report to our knowledge, we will exhibit below a rather
extensive amount of figures obtained through the fully-converged
\pp\ calculation,  but avoid promoting a particular theory for the 
\htc\ superconducting mechanism.

\section{Computational method}
%
%
        We generate soft norm-conserving \pp s using the scheme of
Troullier and Martins~\cite{Troullier}, which gives a very fast
convergence of the total energy with respect to the basis-set size.
(It is possible to use {\em ultrasoft} \pp s~\cite{Vanderbilt} 
which require a much smaller basis set size at the price of 
abandoning the norm-conserving property, but we do not adopt that 
scheme here.)
To enhance the transferability of the \pp\ and guarantee correct
scattering properties of the \pp, we include some semi-core
orbitals into the valence shell and use ionized configurations for
the atoms Y and Ba.  The configuration and the cutoff
radii $r_{cl}$'s with which the \pp s are generated are listed in
\tab{t:PARAMT-PP}.  All the \pp s, except for the oxygen \pp,
are generated semi-relativistically.  The partial core
correction scheme~\cite{Louie} is employed to overcome the problem
associated with the nonlinearity of the exchange-correlation
functional.  The Ceperly-Alder correlation~\cite{Ceperly} is used
in the parametrized version of Perdew and Zunger~\cite{Perdew}.
These \pp s are then cast into the fully-nonlocal separable form of
Kleinman and Bylander~\cite{Kleinman} with use of
$s$-locality to avoid the ghost states~\cite{Gonze90,Gonze91}.
The norm-conserving
requirement results in the matched first energy derivative of the
logarithmic derivative of the wavefunction,~
$\pdp{\pdp{\log{R_l(r)}}{r}}{\epsilon}|_{\epsilon=\epsilon_l,
r=r_{cl}}$,~ between the pseudo radial wavefunction and the
all-electron one, where $R_l(r)$ and $\epsilon_l$ are the radial
wavefunction and the valence orbital energy for angular momentum
$l$.  The \pp s optimized by us with the above scheme show
scattering properties acceptable in a much wider energy range than
enforced by the norm-conserving condition.

   We then perform the plane-wave band structure calculations within 
the LDA\@. The single-particle Kohn-Sham equation is solved by 
diagonalizing the Hamiltonian matrix~\cite{Ihm}. Diagonalization is 
achieved by using the block-Davidson method~\cite{Davidson,Liu} with 
the modified Jacobi relaxation~\cite{Wood,Martins88,Martins91}.
At this point, we want to mention the superiority of the
block-Davidson method with the modified Jacobi relaxation to two 
other competing methods, the \dritz~\cite{Wood,Martins88,Martins91}
and the \pcg~\cite{Teter,Payne} methods, in dealing with systems
with relatively localized states.  We have tested these three 
diagonalization schemes by studying the cubic ZnS with variable 
submatrix sizes.  For relatively small submatrix sizes, the 
block-Davidson method still gives correct results, whereas the 
latter two methods frequently fail to do it.  The discrepancy is 
due to the fact that the block-Davidson method updates the 
eigenvectors simultaneously while the latter two 
adopt a band-by-band updating scheme.  Actually, there exists a 
crossover behavior in relative merits among the three schemes. If the
number of desired states is very large, the computational load in
diagonalizing the Hamiltonian projected to the expansion space is
too severe with the first method and the latter two methods with a
sufficiently large submatrix is preferable to the first one.  As
the number decreases, however, the computational effort of the
first method in the diagonalization of the
expansion-space-projected Hamiltonian is greatly reduced and the
block-Davidson method is superior to the other two.

\section{Results and Discussion}
%
%
  The calculation for the stoichiometric \sybco\ are
performed with use of the recently obtained structural
parameters at 10K~\cite{Sharma}; $a=3.8196\AA$, $b=3.8813\AA$, 
$c=11.64028\AA$, \zcord{Ba} = 0.18388, \zcord{Cu2} = 0.35457,
\zcord{O2} = 0.37823, \zcord{O3} = 0.37813, and \zcord{O4} = 
0.15976.
The atomic positions in lattice unit are Y(\half, \half, \half), 
Ba(\half, \half, $\pm$\zcord{Ba}), Cu1(0, 0, 0), 
Cu2(0, 0, $\pm$\zcord{Cu2}), O1(0, \half, 0), 
O2(\half, 0, $\pm$\zcord{O2}), O3(0, \half, $\pm$\zcord{O3}),
and O4(0, 0, $\pm$\zcord{O4}).  The cutoff energy of \ry{72.25} 
and 147 \k-points in the irreducible Brillouin zone (IBZ, $1/8$ 
of the Brillouin zone) 
with the linear tetrahedron method~\cite{Lehmann,Rath}
are used to obtain the self-consistency.  The self-consistent 
energy bands are plotted in \fig{f:YBCO-band1} along the symmetry 
directions in the IBZ\@. The same band structure is shown in 
expanded scale in \fig{f:YBCO-band2} for the energy range from
\ev{-1.0} to \ev{3.0}\@. There are four partially filled bands
crossing \ef .  Two nearly half-filled bands are derived from the
two \cuo\ planes and have the usual two-dimensional character of
Cu2($3d_{x^2-y^2}$)--O2($2p_x$)--O3($2p_y$) $\sigma$-antibonding.
The other two are derived from the CuO chain and show negligible
dispersions along S-Y and R-T in contrast to appreciable
dispersions along X-S and U-R\@.  The two CuO chain bands show
completely different bonding characters.  The nearly-empty
dispersive band has the character of
Cu1($3d_{y^2-z^2}$)--O1($2p_y$)--O4($2p_z$) $\sigma$-antibonding 
and the almost filled flat band has the
Cu1($3d_{yz}$)--O1($2p_z$)--O4($2p_y$) $\pi$-antibonding character.
Four \ef-crossing bands form a complicated FS's shown in
\fig{f:YBCO-FS}; one stick, two cylinders along the $c$ direction,
and one sheet perpendicular to the $\Gamma$-Y direction.    
   There are many published literatures presenting either the 
energy band structure or the FS's of \sybco, all of 
which resort to localized basis methods. Among them, the first 
report was made by Yu \etal~\cite{Massidda87,JYu87b} and 
illustrates an overall feature of the band structure of \sybco, but 
shows a somewhat unconverged result --- the existence of a hole-like 
cylindrical FS along Y-T\@. Later Massidda~\cite{Massidda90} 
assigned the incorrectness to the small number (10) of 
\k-points and corrected it with use of a larger number (32) of 
\k-points.  The second report by Krakauer
\etal~\cite{Krakauer} shows more converged results than Yu \etal, 
but the interaction between plane bands and the chain band are 
incorrectly described as will be discussed below. 
Results of the subsequent calculation
by the same group (Pickett \etal~\cite{Pickett90}) are 
different from its preceding work; the band-band interaction is 
correctly described but the chain derived band 
along $\Gamma$-Y, which is slightly unfilled in 
Krakauer \etal~\cite{Krakauer}, lies below \ef. Andersen \etal's 
FLMTO results~\cite{Andersen} are indistinguishable from Pickett 
\etal's.  There is a calculation done by using the \pp\ 
mixed-basis method \cite{Hiroi}, but the resulting energy bands 
show poor convergence. Judging from the band structure and the 
calculated FS's, the present \pppw\ calculation shows 
the most converged results with accuracy comparable with Pickett 
\etal~\cite{Pickett90} and Andersen \etal~\cite{Andersen}.
(Though the FS topology is different for the large 
plane-derived band, it is fragile with respect to the variation 
of \ef\ on a scale as small as $\sim$ \ev{0.01}.)
The present results are also in good agreement with
experiments~\cite{Kampuzano,Haghighi,Fowler}.

The bonding nature of the four partially filled bands can be
clarified by examining their charge density contour plots at 
some selected \k-points.  The charge densities of the states 
at {\bf k} = (\quatro, \quatro, 0), a half way from $\Gamma$ 
to S, for the lower and the upper \cuo\ bands are shown in 
\figs{f:YBCO-contour-plane}{a}{b} respectively.  Both of them 
reflect the Cu2($d_\xy$)--O2($p_x$)--O3($p_y$) $\sigma$-antibonding 
hybridization characteristic of the \htc\ cuprates.  In detail, 
however, their symmetries are different. The wavefunction of the 
lower \cuo\ band is the symmetric linear combination of the two 
\cuo\ plane states residing above and below the Y atom plane, 
and the upper \cuo\ band is the antisymmetric one, which can be 
deduced from the comparison of the smallest-value contour lines 
in \figs{f:YBCO-contour-plane}{a}{b}.  
Furthermore, their parity with respect to the chain plane
reflection is even for the lower band and odd for the upper one,
respectively.  Since the wavefunctions of the nearly-empty chain
band are even with respect to the reflection in the plane of the
chains perpendicular to $c$ axis, it follows that for $k_z = 0$ the
chain band does cross the odd (upper) plane band while hybridizing
with the even (lower) plane band.  For $k_z$ = \pioc, the situation
is reversed, \ie, the nearly-empty chain band interacts with the
upper plane band but does not with the lower one. Such 
symmetry properties and the interaction between the plane bands and
the nearly-empty chain band are mentioned in Pickett 
\etal~\cite{Pickett90} and fully described in Mazin 
\etal~\cite{Mazin92b}.

The charge densities of the nearly-empty chain band are shown in
\fig{f:YBCO-contour-chain1} at {\bf k} = (\half, \half, 0) and 
(\quatro, \quatro, 0).  The Cu1($d_\yz$)--O1($p_y$)--O4($p_z$)
$\sigma$-antibonding character and the local four-fold coordination
of the Cu1 atom are clearly identified from them.  The absence of
the band dispersion along S-Y or R-T can be ascribed to the strong
one-dimensional character along the $b$ axis. As the energy rises
towards the S point the antibonding character of the Cu1--O1 bond
increases while that of Cu1--O4 is almost unaffected.  The shorter
Cu1--O4 bond length of $1.8596\AA$ compared with the Cu2--O4
interatomic distance of $2.2676\AA$ is consistent with the bonding
characters identified above.  The Cu1--O1--O4
$pd\sigma$-antibonding band is almost empty, while its bonding
counterpart is completely filled. It is therefore likely for O4
atoms to be tightly bonded to Cu1 atoms. On the other hand, the Cu2
atoms have no ligand O atom on the Y plane side. The Cu2($d_\zr$)
orbitals are only weakly bonded to O4 and both the bonding and
antibonding bands are almost filled, resulting in a weaker Cu2--O4
bond and a longer Cu2--O4 interatomic distance.

The contour plots of the charge density for the fourth (almost
filled) band are shown in \fig{f:YBCO-contour-chain2} at \k\ =
(\half, \half, 0) and (\quatro, \quatro, 0).  A remarkable feature
is that the main contribution to this band comes from O1($p_z$) and
O4($p_y$) orbitals and relatively little charge resides at the Cu,
O2, and O3 atoms.  The character at S is
O1($p_z$)--O4($p_y$)--Cu1($d_{yz}$)--Cu2($d_{yz}$)
$\pi$-antibonding and the charge is depleted on the 
$zx$-plane.  The states near \k\ =(\quatro, \quatro, 0) have the
O1($p_z$), O4($p_x$, $p_y$, $p_z$), Cu1($d_{yz}$, $d_{zx}$),
Cu2($d_\zx$), and O2($p_z$) orbital characters.
Since these chain-related states lie in the narrow energy region
containing \ef, they are expected to be strongly affected as the
oxygen content on the CuO chain is reduced and the hole doping
concentration decreases, implying that \Tc\ should be influenced
significantly by these states.

The muffin-tin radii of 2.80, 3.20, 1.95, and \au{1.65} for Y, Ba,
Cu, and O atoms respectively are used in calculating the partial
charges and the partial (or site-projected and symmetry-decomposed)
densities of states (PDOS)\@. The total density of states (TDOS) at 
\ef\ is \cudos{1.41}.  For comparison, Massidda 
\etal~\cite{Massidda87} obtained \nef = \cudos{1.13} and Krakauer 
\etal~\cite{Krakauer} 1.38.
The PDOS plots in
\fig{f:YBCO-dos1} show that the Cu1--O1--O4 sates prevail in the
energy range from \ev{0.5} below \ef\ to \ef.  The PDOS in an
expanded scale in \fig{f:YBCO-dos2} indicates a van Hove 
singularity at $\sim$ \ev{0.03} below \ef\ with a clear Cu1--O1--O4
hybridization.  The importance of the van Hove singularity to the 
\htc\ mechanism has been strongly advocated in some theoretical
studies~\cite{Tsuei}.  The partial charges and their symmetry
decompositions are listed in \tab{t:PC-ybco}. 
(The partial charges and their symmetry decompositions are presented 
and fully analyzed in Schwarz \etal~\cite{Schwarz}.)
The relatively small amount of O1($p_y$), O4($p_z$), Cu1($d_\xy$), 
and Cu1($d_\zr$) partial charges are due to the nearly-empty 
dispersive band of Cu1($d_\yz$)--O1($p_y$)--O4($p_z$) hybridization. 
In fact, with use of a different choice of symmetry decomposition 
including $d_{y^2-z^2}$ symmetry, the partial charge of 
Cu1($d_{y^2-z^2}$) is further reduced to 1.30, which is smaller than 
the partial charge of Cu1($d_\zr$), 1.38.  The relatively small 
amount of O2($p_x$), O3($p_y$), and Cu2($d_\xy$) partial charges 
compared with the other O2($p$), O3($p$), and Cu2($d$) partial 
charges is attributed to the characteristic $pd\sigma$-antibonding 
hybridization [Cu2($d_\xy$)--O2($p_x$)--O3($p_y$)] of the
superconducting cuprates. The smaller partial charge of
Cu1($3d_{y^2-z^2}$) than that of Cu2($3d_{x^2-y^2}$) (1.30 versus
1.42) is in accordance with the existence of the nearly-empty chain
band and two nearly half-filled plane bands.  The partial charges
of the Cu($4p$) states are relatively small, but their magnitude
can serve as a measure of the bond strength between a copper atom
and its neighboring atoms.  The Cu($4p$) partial charges do not
originate from {\em on-site} Cu($4p$) states, which lie at much
higher energies, but represent the intrusion of wavefunctions
centered at neighboring (off-site) atoms into the copper
muffin-tin sphere.  They originate mainly from the surrounding
O($2p$) wavefunctions and their magnitude depends on the distance
from copper to the neighboring oxygen sites.  Since the Cu1--O4
bond length is shorter than that of Cu1--O1, the partial charge of
Cu1($p_z$) is larger than that of Cu1($p_y$). The Cu1($p_x$)
partial charge is the smallest due to the absence of the oxygen
neighbors in the $a$ direction.  For the plane copper atoms, the
longer distance to O4 than to O2 or O3 manifests itself in the
smaller Cu2($p_z$) partial charge than that of Cu2($p_x$) or
Cu2($p_y$).

\section{Summary}
%
%
   We have applied, for the first time, the \abi\ \pppw\ method
to the electronic structure calculation of the \htc\ 
cuprate \sybco\ within the LDF formalism.
Soft \pp s of optimized transferability are generated through the
Troullier-Martins scheme and cast into the Kleinman-Bylander type
fully-separable form.  The block-Davidson algorithm with the
modified Jacobi relaxation operator is then used to solve the
single-particle Kohn-Sham equation.  
The energy bands, the Fermi surfaces, the charge densities, the total 
DOS, the partial DOS, and the symmetry-decomposed partial charges 
obtained here are in good agreement with both the
existing theoretical calculations using (F)LMTO and (F)LAPW methods 
and the available experimental facts.  
This successful application to \sybco\ has demonstrated 
that one can deal with systems containing first row elements, 
transition metals, or rare-earth elements 
with acceptable computational efforts via the \abi\ \pppw\ method.

Helpful discussions with Prof.~Jaejun Yu are greatly appreciated.
This work is supported by the Ministry of Science and Technology,
the '94 Basic Science Research Program of the Ministry of
Education, the SNU Daewoo Research Fund, and the Korea Science and 
Engineering Foundation through the SRC program.

%
%

\begin{table}
\caption{The reference configurations and core radii.
         For Y and Ba atoms, $+3$ and $+2$ 
         ionized configurations are used respectively to get optimal
	 pseudopotentials.}
\vskip 0.2truecm
\begin{tabular}{c|cccccc}
 atom &  Y   &   Ba  &   Cu   &   O   \\
 \hline
configuration & 4$s^2$4$p^6$4$d^1$ &  5$s^2$5$p^6$5$d^0$ 
& 4$s^1$4$p^0$3$d^{10}$ & 2$s^2$2$p^4$ \\
 \hline
 $r_{cs}$ & 1.19 & 1.65 & 2.06 & 1.15 \\
 $r_{cp}$ & 1.35 & 2.00 & 2.30 & 1.45 \\
 $r_{cd}$ & 2.49 & 2.48 & 2.06 &   -  \\
\end{tabular}\label{t:PARAMT-PP}
\end{table}
\begin{table}
\caption{The symmetry-decomposed partial charges (in electrons) 
         of the valence states of \sybco.}
\vskip 0.2truecm
\begin{tabular}{l|c|cccc|cccccc|c}
 atom &  $s$ &  $x$ &  $y$ &  $z$ &  $p$ & $xy$ & $yz$ & $zx$ & 
$\xy$ & $\zr$&  $d$ & total \\
 \hline
  Y   & 0.086 & 0.069 & 0.066 & 0.070 & 0.204 & 0.048 & 0.136 & 
0.146 &    0.137   &   0.052   & 0.521 &  0.811 \\
  Ba  & 0.062 & 0.135 & 0.123 & 0.120 & 0.377 & 0.092 & 0.039 & 
0.058 &    0.037   &   0.060   & 0.286 &  0.725 \\
 Cu1  & 0.194 & 0.028 & 0.059 & 0.079 & 0.167 & 1.836 & 1.856 & 
1.837 &    1.644   &   1.379   & 8.552 &  8.912 \\
 Cu2  & 0.175 & 0.063 & 0.059 & 0.033 & 0.156 & 1.848 & 1.811 & 
1.809 &    1.419   &   1.738   & 8.627 &  8.957 \\
  O1  & 0.039 & 1.238 & 0.984 & 1.300 & 3.522 & 0.003 & 0.003 & 
0.000 &    0.003   &   0.001   & 0.009 &  3.570 \\
  O2  & 0.043 & 1.068 & 1.287 & 1.251 & 3.606 & 0.003 & 0.000 & 
0.003 &    0.004   &   0.001   & 0.011 &  3.660 \\
  O3  & 0.042 & 1.285 & 1.064 & 1.252 & 3.600 & 0.003 & 0.003 & 
0.000 &    0.003   &   0.001   & 0.010 &  3.652 \\
  O4  & 0.041 & 1.228 & 1.258 & 1.041 & 3.527 & 0.000 & 0.002 & 
0.002 &    0.000   &   0.004   & 0.008 &  3.576 \\
\end{tabular}\label{t:PC-ybco}
\end{table}

\begin{center}
  {\Large\bf Figure Captions}
\end{center}
\newcounter{fignum}
\begin{list}%
{FIG.~\arabic{fignum}.}{\usecounter{fignum}
                       \setlength{\rightmargin}{\leftmargin}}

  \item   The energy band structure of \sybco.
	       \label{f:YBCO-band1}

  \item   The energy band structure of \sybco\
          on  (a) $k_z = 0$ plane
          and (b) $k_z$ = \pioc\ plane in expanded scale.
	       The inset in (b) is the band structure in much more 
          expanded scale illustrating the band crossing between 
          the chain band and the lower plane band as well as the
          band repulsion between the chain band and 
	       the upper plane band.
          \label{f:YBCO-band2}

  \item   The FS's of \sybco\ derived
          from the four partially filled bands. The solid line is the
          stick-like FS along S-R from the almost 
          filled chain band, 
          the short-dashed line indicates the one-dimensional FS
          arising from the nearly-empty chain band,
          the long-dashed line is the cylindrical FS from
          the upper \cuo\ band, and the medium-dashed line from the 
          lower \cuo\ band.
          \label{f:YBCO-FS}

  \item   The charge density contours of \sybco\ for states at
          \k\ = (\quatro, \quatro, 0) of 
          (a) the lower Cu2--O2--O3 plane band 
          and (b) the upper Cu2--O2--O3 plane band.
          Contour values run from 0.0002 to 0.0102 in steps of 
          0.002~e/unit-cell.
          The symbols \Bullet\ and \solidiamond\
          represent Cu and O atoms respectively.
          \label{f:YBCO-contour-plane}
  
  \item   The charge density contours of \sybco\ for the 
          nearly-empty Cu1--O1--O4 $\sigma$-antibonding chain 
          band at (a) \k\ = (\half, \half, 0) and
          (b) \k\ = (\quatro, \quatro, 0).
          Contour values run from 0.001 to 0.011 in steps of 
          0.002~e/unit-cell.
          \label{f:YBCO-contour-chain1}

  \item   The charge density plots of the almost filled
          chain-related band of \sybco\ at
          (a) \k\ = (\half, \half, 0) and
          (b) \k\ = (\quatro, \quatro, 0).
          Contour values are the same as in 
          \fig{f:YBCO-contour-chain1}.
          \label{f:YBCO-contour-chain2}

  \item    The total and the partial densities of states of 
           \sybco . (a) The TDOS, (b) the PDOS of chain-related 
			  atoms, and (c) the PDOS of plane-related atoms.
           \label{f:YBCO-dos1}

  \item    (a) The TDOS, (b) the PDOS of chain-related atoms, and
	        (c) the PDOS of plane-related atoms
           of \sybco\ near \ef\ in expanded scale.
           \label{f:YBCO-dos2}
\end{list}

\end{document}